 \documentclass[12pt,preprint]{aastex}
 \usepackage{emulateapj5}
 \usepackage{apjfonts}

\def\myputfigure#1#2#3#4#5%
{\vskip#5pt\makebox[0pt]{\hskip#2in
\includegraphics[width=#3\textwidth]{#1}}\vskip#4pt\hfill}

\newcommand\lsim{\mathrel{\rlap{\lower4pt\hbox{\hskip1pt$\sim$}}
        \raise1pt\hbox{$<$}}}
\newcommand\gsim{\mathrel{\rlap{\lower4pt\hbox{\hskip1pt$\sim$}}
        \raise1pt\hbox{$>$}}}
\shorttitle{Reionization from decaying particles}
\shortauthors{Hansen \& Haiman}
\begin{document}
\title{%
Do We Need Stars to Reionize the Universe at High Redshifts?\\
Early Reionization by Decaying Heavy Sterile Neutrinos
}%

\author{Steen H. Hansen$^1$ \& Zolt\'an Haiman$^2$}
\affil{%
$^1$ University of Zurich, Winterthurerstrasse 190,
8057 Zurich, Switzerland\\
$^2$ Department of Astronomy, Columbia University, New York, NY 10027, USA
}%

\begin{abstract}
A remarkable result of the Wilkinson Microwave Anisotropy Probe ({\it
WMAP}) observations is that the universe was significantly reionized
at large redshifts. The standard explanation is that massive stars
formed early and reionized the universe around redshift $z\approx
17$. Here we explore an alternative possibility, in which the universe
was reionized in two steps. An early boost of reionization is provided
by a decaying sterile neutrino, whose decay products, relativistic
electrons, result in partial ionization of the smooth gas.  We
demonstrate that a neutrino with a mass of $m_\nu\sim 200$ MeV and a
decay time of $t\sim 4\times 10^{15}$s can account for the electron
scattering optical depth $\tau\approx 0.16$ measured by {\it WMAP}
without violating existing astrophysical limits on the cosmic
microwave and gamma ray backgrounds.  Reionization is then completed
by subsequent star formation at lower redshifts. This scenario
alleviates constraints on structure formation models with reduced
small-scale power, such as those with a running or tilted scalar
index, or warm dark matter models.
\end{abstract}

\section{Introduction}
\label{introduction}

The recent observations by the Wilkinson Microwave Anisotropy Probe
({\it WMAP}) have clarified many issues related to the energy content
of the universe, and determined key cosmological parameters to
unprecedented precision~\citep{spergel03}.  One of the interesting
results of {\it WMAP} comes from the cross-correlation between the
temperature and E--mode polarization angular power spectra, which has
shown that the electron scattering optical depth is large, $\tau_e =
0.16 \pm 0.04$~\citep{kogut03}. This high value can be explained by an
early phase of star--formation, reionizing the universe in a single
step at redshift $z=17\pm0.3$.  A series of papers following the {\it
WMAP} results have described this scenario in detail (Wyithe \& Loeb
2003b; Haiman \& Holder 2003; Ciardi et al. 2003; Somerville \& Livio
2003; Fukugita \& Kawasaki 2003; Sokasian et al. 2003; Cen 2003b; Chiu
et al. 2003).  The implications of the {\it WMAP} results for the
stellar reionization scenario has been reviewed recently in Haiman
(2003).

In structure formation models in the currently favoured spatially flat
cosmologies with dark energy and cold dark matter ($\Lambda$CDM), the
earliest large dark matter structures appear at redshift $z\sim 20$. Stellar
reionization is therefore quite natural, provided that these early
structures are associated with sites of efficient star formation and
ionizing photon production.  Nevertheless, an optical depth as high as
$\tau_e = 0.16$ requires efficiencies that are higher than expected
from the known population of galaxies at lower redshifts.  Most
pre-{\it WMAP} studies of cosmological reionization favoured a
reionization epoch around $z=7-12$ (Haiman \& Loeb 1997, 1998; Gnedin
\& Ostriker 1997), although significant reionization at higher
redshifts was predicted in models that included an early generation of
massive, metal--free stars (Cen 2003a; Wyithe \& Loeb 2003a).

Previous evidence based on the spectra of distant quasars have shown
that the intergalactic medium (IGM) is highly ionized at least out to
redshifts $z\sim 6$. However, the most natural interpretation of the
strong HI absorption seen in the spectra of the highest redshift
quasars (Becker et al. 2001; Fan et al. 2003) in the Sloan Digital Sky
Survey (SDSS) is that reionization is being completed near the
redshifts of these sources, at $z\sim 6-7$.  The {\it thermal} history
of the IGM contains additional evidence that reionization was
completed at $z<10$ (Hui \& Haiman 2003; Theuns et al. 2002).

Post-{\it WMAP} analyses (e.g. Haiman \& Holder 2003; Cen 2003b; Wyithe
\& Loeb 2003b) indicate that such an early generation of metal--free
stars, or an equivalent 'boost' by a factor of $\gsim 20$ in the
efficiency parameters of the $z\sim 20$ ionizing source population, is
required relative to that of the $z\sim 6$ source population, in order
to {\it simultaneously} match the {\it WMAP} value of $\tau_e=0.16$
and the SDSS quasar data.  It is worth noting that stellar
reionization would be much less realistic in cosmologies with low
small-scale power, in which the onset of structure formation is
delayed, such as those with a running or tilted scalar index
(Hannestad et al. 2002; Haiman \& Holder 2003; Somerville et
al. 2003), or warm dark matter models (Barkana, Haiman \& Ostriker
2001; Yoshida et al. 2003; Somerville et al. 2003b; Spergel et
al. 2003).

In summary, in order to explain the large value of $\tau_e=0.16$, the
standard stellar reionization scenario requires {\it 1)} a hitherto
unobserved early stellar population, and {\it 2)} small--scale dark
matter structure to be in place by $z\sim 20$.  While these
requirements are reasonable, whether they are satisfied remains to be
tested with future data. In this paper, we pose the question:
{\it are there alternative possibilities to significantly reionize the
universe at $z\sim 20$?}  One such alternative that avoids the above
requirements is reionization by decaying particles.  \citet{sciama82}
suggested the possibility of photo-reionization from radiatively
decaying particles where the decaying particle has a mass of few tens
of eV.  Related issues were discussed thoroughly in numerous
papers~\citep[early studies
include][]{Cabibbo:er,salati84,rephaeli81}.  However, these types of
particle physics candidates, 
such as light
photinos and gravitinos, are now excluded by modern accelerator
data~\citep{Hooper:2002nq,dreiner03}.  Another interesting candidate was a
decaying active neutrino~\citep{sciama90} which was proposed both as a
dark matter candidate and also to reionize the universe. However,
current cosmological data excludes such massive
neutrinos~\citep{sarkar98,bowyer,spergel03}.

Let us imagine the general case of a cosmologically significant
particle, with the property that it decays around a redshift $z\sim
20$. The decay products could have energies high enough to reionize
the light elements.  If this hypothetical particle is sufficiently
abundant, then it may explain the high value of the optical depth,
without the need for stars to reionize the universe early.  In such a
scenario, normal stellar populations could form at lower redshifts
($z\lsim 10$), with the usual efficiencies, seeded in dark matter
structures on the better understood scales of dwarf galaxies, and
account for the completion of reionization epoch that appears to be
occurring at $z<10$.

In this paper, we turn to particle physics and describe a
specific example for such a particle that is both viable (on particle
physics grounds) and satisfies existing astronomical constraints.  In
practice, the choice of a specific particle candidate turns out to be
severely limited by astronomical observations, such as big bang
nucleosynthesis (BBN), the diffuse gamma--ray background (DGB), and
the stringent limit on the deviation of the spectrum of the cosmic
microwave background (CMB) from a pure Planck shape.  Here we present
a model which relies on a minor extension of the standard model:
introducing a heavy sterile neutrino which mixes with one of the
active neutrinos.  Heavy sterile neutrinos are expected in extensions
of the standard model, in order to give masses to the active (light)
neutrinos through the see--saw mechanism~\citep{yanagida,mohapatra}.
The neutrino masses we will consider below are somewhat smaller than
invoked for the see--saw mechanism.  Our particle should therefore be
regarded as a new independent sterile neutrino \citep[see e.g.][for
reviews on heavy sterile neutrinos in cosmology]{hansen01,dolgov03}.
For the purpose of reionization, the main novelty of this particle is
that its decay products are energetic electrons, rather than photons.
As we shall see below, this makes it possible to achieve reionization
without violating other astrophysical observations.

The rest of this paper is organized as follows. In
\S~\ref{decaying}, we describe the candidate particle we propose,
quantify the expected amount of reionization, and show that the
particle obeys various astrophysical constraints. In
\S~\ref{discussion}, we discuss the cosmological implications of this
new particle, and in \S~\ref{conclusions}, we offer our conclusions.
Throughout this paper, we adopt the background cosmological parameters
as measured by the {\it WMAP} experiment, $\Omega_m=0.29$,
$\Omega_{\Lambda}=0.71$, $\Omega_b=0.047$, $h=0.72$ (Spergel et
al. 2003, Tables 1 and 2).

\section{Decaying sterile neutrino}
\label{decaying}

\subsection{Production and Decay Mechanisms}

We will now discuss some details of the production and decay mechanism
of the heavy sterile neutrino we propose. Our main goal is to
demonstrate that particle physics can provide a viable decaying
particle whose decay products (energetic electrons in our case) can
reionize the universe early, without violating other observations.  We
do not attempt a comprehensive analysis of all possible particle
physics candidates here; indeed, we expect that there could be other
particle candidates which could similarly account for early
reionization.

Let us consider a standard two-neutrino oscillation scheme, where one
of the active neutrinos, $\nu_\alpha = \nu_e$, $\nu_\mu$ or
$\nu_\tau$, mixes with a heavy mainly sterile neutrino, $\nu_s$,
\begin{eqnarray} 
\nu_\alpha &=& \cos\theta ~\nu_1 + \sin
\theta ~\nu_2\,, \nonumber \\ 
\nu_s &=& -\sin \theta ~\nu_1 +
\cos\theta ~\nu_2\,,
\label{eq:mixing}
\end{eqnarray} 
where $\nu_1$ and $\nu_2$ are assumed to be the light and heavy mass
eigenstates respectively, and $\theta$ is the vacuum mixing angle.
Let the light neutrino be massless and the heavy neutrino have mass
$m$.  We will consider very small mixing angles, and hence refer to
the light neutrino mass eigenstate as the active neutrino and the
heavy one as the sterile neutrino.

If the sterile neutrino has a mass in the range 140-500 MeV, then the
dominant decay channel~\citep{Astier:2001ck} is
\begin{equation}
\nu_s \rightarrow l + \pi \, ,
\end{equation}
where the lepton, $l$, can be either an electron, a positron, or a
neutrino.  For larger masses, new decay channels open up (including
kaons, eta, etc.), and the considerations below become more
complicated. For smaller masses the decay products include too few
photons and electrons for our purpose.
The rate for the reaction $\pi_0 \rightarrow \nu \bar
\nu$ was calculated by~Fischbach et al. (1977) (see also
Kalogeropoulos et al. 1979), and the result was converted to the
reaction $\nu_s \rightarrow \nu_\alpha + \pi_0$ by \citet{dhrs00}.
Here we also need to include the charge current reactions
\begin{eqnarray}
\nu_s &\rightarrow& e^{-} + \pi_+ \, , \nonumber \\
\nu_s &\rightarrow& e^{+} + \pi_- \, ,
\label{eq:epi}
\end{eqnarray}
and we find the lifetime
\begin{eqnarray}
\tau_{\rm decay} &=& \left[ \left(2+r^2\right) 
\frac{(G_F f_\pi {\rm sin}\theta)^2}{8\pi}  
m\left(m^2-m_\pi^2 \right)\right]^{-1}\nonumber \\
&\approx&  \frac{1.2 \times 10^{-9} {\rm sec}}{M(M^2 -1)
{\rm sin}^2\theta} \, ,
\end{eqnarray}
where $r^2 \approx 0.2$, $G_F$ is the Fermi constant, $f_\pi=131$ MeV,
and the masses $M = m/m_\pi$ are all in units of the charged pion mass,
$m_{\pi}=139.7$ MeV.  When we express the decay time,
$\tau$, in units $10^{15}$ sec $\approx 30$ Myr, then this equation
becomes
\begin{equation}
\tau_{15}  \, M(M^2 -1) {\rm sin}^2\theta = 1.2 \times 10^{-24} \, .
\label{eq:decay}
\end{equation}
The reactions~(\ref{eq:epi}) directly provide us with a significant
number of highly relativistic electrons and positrons, which we can
use to reionize the universe early. Some of the energetic positrons
could annihilate with the thermal (and also with the much more dilute,
energetic) electrons, but the resulting photons would have mean free
path larger than the Hubble distance. Such annihilation would change
our result slightly at worst, by reducing the number of
ionizations. 

Let us now turn to the production and abundance of the sterile
neutrinos.  These neutrinos are produced through standard neutrino
oscillations while they are still relativistic, at a temperature of
about 7GeV, when there are about 90 degrees of freedom
in the plasma~\citep{kolbturner}.  For
mixing with electron neutrinos, the fraction of sterile neutrinos to
active ones is~\citep{dh01}
\begin{equation}
\frac{n_s}{n_\alpha} = 4.0 \times 10^{9} \, {\rm sin}^2\theta \, M \, .
\label{eq:production}
\end{equation}
For mixing with muon or tau neutrinos, the numerical coefficient in front 
is instead $5.6$. 
If the sterile state is mixed with both electron and muonic (tau)
neutrinos, then the produced number increases correspondingly.  We
thus assume that there is no thermal relic abundance of these sterile
neutrinos. Specific models with such properties exist
\citep{bento,berez}, however, other production mechanisms can produce
a significantly higher abundance of sterile
neutrinos~\citep{Sarkar:1995dd,mohapatra98}.

\subsection{Reionization Efficiency}

Let us next estimate the amount of hydrogen which can be reionized by
the decay products.  The universe is transparent to the vast majority
of photons from the process $\pi_0 \rightarrow 2 \gamma$, since these
photons have energies above $m_\pi/2$.  
Since the abundance of these sterile neutrinos is about 15 orders of
magnitude below that of a thermalized neutrino, one easily
sees~\citep{kolbturner} that the diffuse gamma background is many
orders of magnitude larger than the photons resulting from both
$\pi_0$ and $\pi^\pm$.  We can therefore ignore this fraction of the
decay products and consider the fate of the relativistic decay
electrons instead. These electrons are injected
into the IGM at redshift $z\sim 20$ (i.e. at the redshift
where the decay time equals the age of the universe) 
with energies of about $E_e =
(m_\pi/2)(M-1)$, i.e. between $0 < E_e < 180$ MeV.  The relevant
interactions for such electrons with the background gas and with the
CMB have been summarized in, e.g. Oh (2001).  For our purposes, the
two important processes are inverse Compton scattering with CMB
photons, and collisional ionization of neutral hydrogen.  The mean
free path to Compton scattering is 
\begin{equation}
\lambda_C= (\sigma_T
n_\gamma)^{-1}\sim 4\times10^{17}[(1+z)/21]^{-3} {\rm cm} \, , 
\end{equation}
where $\sigma_T$
is the Thomson cross section and $n_\gamma$ is the CMB photon number
density. The mean free path to collisionally ionizing hydrogen is
\begin{equation}
\lambda_H= (\sigma_H n_H)^{-1}\sim 1.4\times10^{24} (E_e/100{\rm
MeV})[(1+z)/21]^{-3} {\rm cm} \, ,
\end{equation} 
where $\sigma_H$ is the electron impact
ionization cross section (Shull \& van Steenberg 1985, equation A2; we
have ignored a small logarithmic correction), and $n_H=2\times10^{-3}
[(1+z)/21]^{3}\,{\rm cm^{-3}}$ is the neutral hydrogen density.  Since $\lambda_C\ll
\lambda_H$, the electrons will undergo repeated Compton scatterings
with the CMB photons.  At each scattering event, the CMB photon will
be up--scattered to the energy $E_\gamma \sim (4/3)\gamma^2
E_{\gamma,0} \sim 300 (E_e/100{\rm MeV})^2 [(1+z)/21]$ eV, where
$0<\gamma<360$ is the Lorentz factor for the electron, and
$E_{\gamma,0}=5\times10^{-3} [(1+z)/21] $ eV is the mean (undisturbed)
CMB photon energy. 
The up-scattering of the CMB photons is the {\it first step} (of 3 steps)
towards reionization.
The electron will loose the same amount of energy
$E_\gamma$, and will cool after $N\sim E_e/E_\gamma= 3\times 10^5
(E_e/100{\rm MeV})^{-1}[(1+z)/21]^{-1}$ scatterings, corresponding to
the cooling time $\tau_C\sim N\lambda_C/c\sim 6\times10^{4}
(E_e/100{\rm MeV})^{-1} [(1+z)/21]^{-4}$ yr.  The cooling time for the
electron due to collisionally ionizing hydrogen, on the other hand, is
$\tau_{\rm H}\sim 2.9\times10^{9} (E_e/100{\rm MeV})^{-1}
[(1+z)/21]^{-3}$ yr.

By equating $\tau_C= \tau_H$ we infer that decay electrons whose
energies are above $E_e> 1 [(1+z)/21]^{-1/2}$ MeV will loose most of
their energy to up--scattering the CMB photons.  We conclude that only
electrons that are produced by sterile neutrinos very near (within
$\sim 2$ MeV) the pion mass could spend a significant fraction of
their initial energy on direct collisional H-ionizations.  However,
CMB photons that are up--scattered by electrons with energies $20
[(1+z)/21]^{-1/2}\, {\rm MeV} < E_e < 180$ MeV will have final
energies 
\begin{equation}
13.6 \, {\rm eV} < E_\gamma < 900[(1+z)/21] {\rm eV} \, .
\end{equation}
These
photons then spend a significant fraction of their energy on
reionizing hydrogen. At the low (UV) end of this energy range, the
Comptonized photons directly photoionize the neutral hydrogen atoms.
Despite the steep decrease of the photo-ionization cross--section
($\sigma_\nu\propto\nu^{-3}$), even at the high (soft X--ray) end of
the range, the mean free path of the photons to photo-ionization is a
small fraction $ (c \sigma_\nu n_H t_{\rm Hubble})^{-1} \sim 0.2
(E_e/180{\rm MeV})^{3} [(1+z)/21]^{3/2}$ of the Hubble distance.  As a
result, each photon ionizes a hydrogen atom, and creates a fast
secondary photoelectron.  
The photo-ionization and production of the energetic photoelectron
is the {\it second step} towards reionization.
At these low (non--relativistic) electron
energies, inverse Compton scattering is negligible compared to
collisional ionization, and a fraction $\sim 1/3$ of the initial
energy of the photoelectron is subsequently spent on ionizing hydrogen
(Shull \& van Steenberg 1985; Oh 2001; Venkatesan et al. 2001).
This collisional ionization is the {\it third step} towards reionization.

We conclude that a significant fraction ($\gsim 1/3$) of the decay
electrons can produce hydrogen ionizations.  The maximal number of
hydrogen atoms that could be ionized by a single electron, on
energetic grounds alone, would be $\chi=E_e/13.6$eV. This can be
written as $\chi = 5.1 \times \, 10^6 (M-1)$, where $M$ is again in
units of the pion mass.  Thus the number of reionized hydrogen atoms
is given by
\begin{equation}
n_b = \epsilon \chi n_e^{\rm prod} \, ,
\end{equation}
where we have introduced an efficiency factor, $\epsilon \leq 1$, and
$n_e^{\rm prod}$ is the number density of electrons (plus positrons)
from reaction~(\ref{eq:epi}).  Using $n_b/n_\gamma = 6 \times
10^{-10}$ and $n_\alpha = 3/11 \, n_\gamma$, we find that the
required efficiency is
\begin{equation}
\epsilon = 0.10 \, \tau_{15} \left( M +1 \right) \, .
\end{equation}
For mixing with muon (or tau) neutrinos the factor in front is
$0.071$.  Note that this efficiency is quite reasonable: using
$\epsilon =1/3$ (see above discussion) and assuming mixing with an
electronic neutrino, one can reionize half of the baryons with $\tau_{15}
(M+1) = 6.7$ (see Fig. 1).

\vspace{-6\baselineskip}
\myputfigure{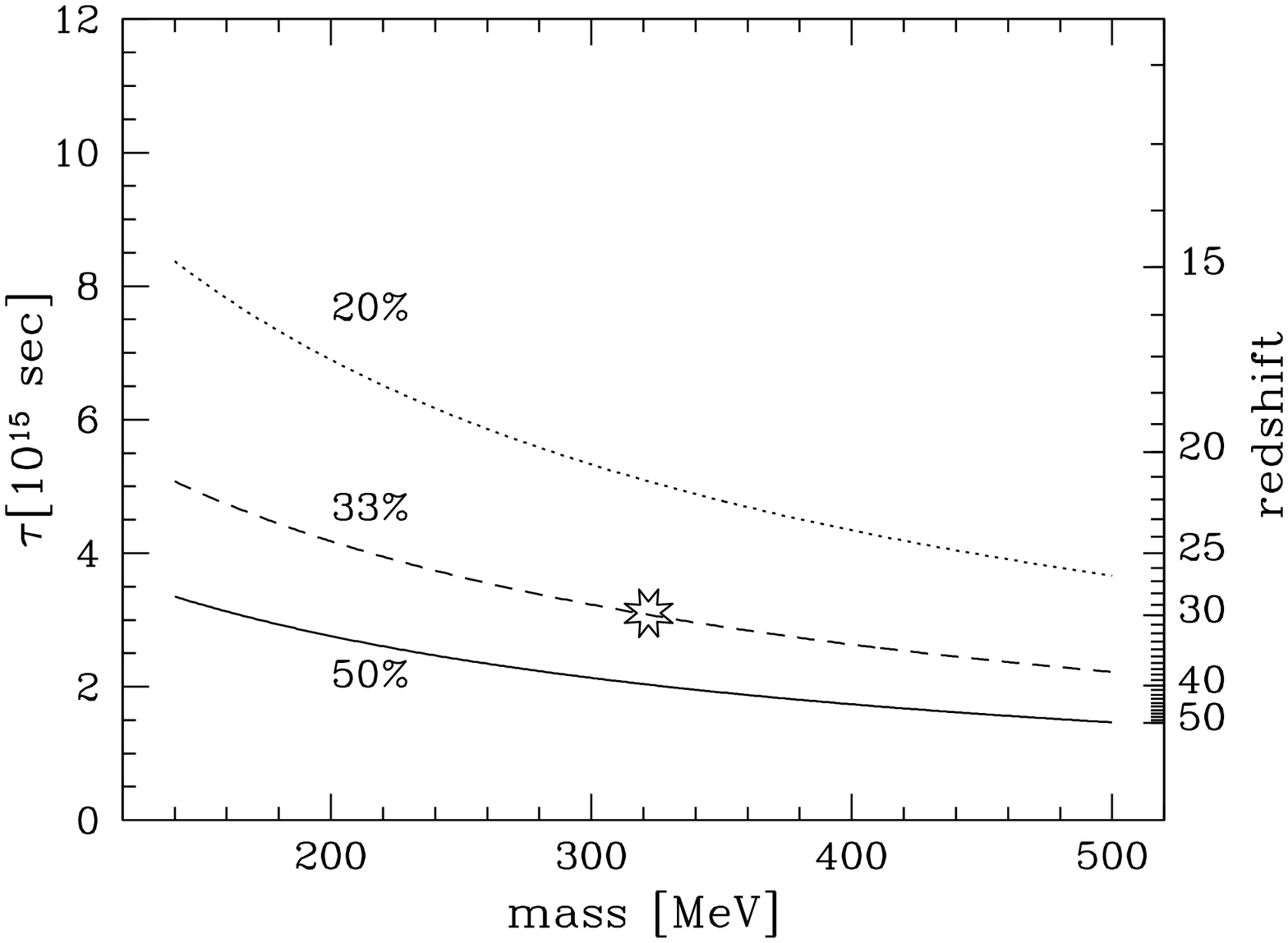}{3.3}{0.5}{.}{0.}  
\vspace{-1\baselineskip}
\figcaption{Parameter space
(decay time vs. mass) for a decaying sterile neutrino mixed with a
electron neutrino, using a reionization efficiency $\epsilon = 1/3$. 
The right axis shows the redshift corresponding to the given decay time.
The solid curve corresponds to ionizing half of the baryons, while the
dashed and dotted curves correspond to ionizing a smaller fraction of
1/3 and 1/5, respectively. The star indicates the life-time
corresponding to the redshift where the universe should be reionized
to the fraction $33\%$ in order to give an optical depth of
$0.16$.}


\subsection{Astrophysical Constraints}

Let us next turn toward the observational consequences of the decaying
sterile neutrinos.  First, we can consider how many sterile neutrinos
were produced, and whether they will affect the production of light
elements (BBN), photon decoupling, or supernova cooling. Using
equations~(\ref{eq:decay}) and (\ref{eq:production}), we find
\begin{equation}
\frac{n_s}{n_\alpha} \approx 5 \times 10^{-15} \, \frac{1}{\tau_{15}\left( 
M^2 -1 \right)}  \, .
\end{equation}
For masses in the range considered ($140$ MeV$<m<500$ MeV) and for a
lifetime several orders around $\tau = 10^{15}$ sec, this is such a
small number that none of the three processes above will be affected
in any way~\citep{Kainulainen91,Sarkar:1995dd,Dolgov:2000pj,dolgov03}.
The neutral pions decay immediately into two energetic photons;
however, the observed diffuse gamma ray
background is several orders of magnitude higher than the 
signal from the decay products~\citep{kolbturner}.  Furthermore, the
mixing angle is so small (see eq.~\ref{eq:decay}), that no terrestrial
experiment can detect the oscillations.

A potential observable from reionization at high redshift is the {\it
spectral} distortion of the CMB.  The CMB photons have been observed
by the {\it COBE} satellite to obey a nearly perfect Planck
distribution. The {\it COBE} results translate into an upper bound on
the energy deposited by the electrons, usually expressed in terms of
the limits $\mid y \mid < 1.5\times 10^{-5}$ and $\mid \mu \mid <
9\times 10^{-5}$ on the Compton $y$ parameter and the dimensionless
chemical potential, respectively (Fixsen \& Mather 2002; Fixsen et
al. 1996). As we showed above, in our scenario, a significant fraction
of the total energy of the decay electrons is transfered to the CMB.
For a sterile neutrino near the pion mass threshold ($1<M<1.02$), for
which the decay electrons collisionally ionize (and heat) the IGM
directly, this transfer occurs by the non--relativistic Compton
coupling between the reionized IGM and the CMB. For high energy decay
electrons, the transfer occurs in two steps. CMB photons are first
up--scattered to UV or soft X--ray energies, effectively removing a
small fraction of the photons from the COBE range, corresponding to a
$\mu$-distortion.  Second, this energy is transfered back to the COBE
range by non--relativistic coupling between the reheated IGM and the
CMB (corresponding to a $y$--distortion).

We therefore compute the total energy in the primary decay electrons
(plus positrons), $\rho_e$, and compare this to the total energy of
the CMB, $\rho_\gamma$. Using the CMB temperature $T\approx 5 \times
10^{-3} eV$ (at redshift z=20) we find $\rho_e/\rho_\gamma \approx 7
\times 10^{-6} /(\tau_{15} (M+1))$.  Depending on the actual fraction
(0.1-0.9) of the mean electron energy that goes into heating the IGM
(rather than going into the ionizations and excitations), the number
can be even smaller. We conclude that both the $\mu$-- and
$y$--distortions are sufficiently small (both $< \rho_e/\rho_\gamma$)
that there is no disagreement with the current upper limit from {\it
COBE}.  However, the decaying neutrino model presented here should
predict a characteristic shape for the distorted CMB spectrum that
could be detected in future measurements improving the sensitivity by
nearly two orders of magnitude (Fixsen \& Mather 2002).  Finally, we
note that in the stellar reionization case, where the ionizing sources
are strongly clustered, the angular {\it fluctuations} of the spectral
distortion can place interesting limits on models (Oh et al. 2003).
Since the electrons from the decaying neutrinos are smoothly
distributed in space, our scenario avoids this constraint by
construction.

\section{Discussion}
\label{discussion}

Let us briefly summarize the scenario envisioned in this
paper. Sterile neutrinos with masses of about a few hundred MeV were
created in the early Universe with an abundance about five orders of
magnitude below the number density of baryons. A large fraction,
$20-30\%$ of the neutrino's mass results in decay--produced energetic
electrons and positrons.  Depending on their energies, these electrons
then produce a UV/soft X-ray background by inverse Compton scattering
CMB photons.  The newly created energetic photons then ionize the IGM
either by direct photo-ionization, or by the collisions between the
fast secondary photoelectrons and the neutral hydrogen atoms.  For a
sterile neutrino mass near (within $\sim 2$ MeV) the pion mass
threshold, the ionization is by direct electron--hydrogen collisions.
As a result, any sterile neutrino with a mass between $140 < M < 500$
MeV would produce relativistic electrons that spend $\gsim 1/3$ of
their energy on partially reionizing the universe.

In this scenario, the sterile neutrinos produce electrons at a rate
$\propto\exp(-t/\tau)$.  For sterile neutrino masses near the lower
mass ranges considered here, which result in electrons with $E\lsim
100$ MeV and ultimately produce UV photons, the ionization time is
always shorter than the Hubble time, including at redshift $z\sim
1000$.  As a result, reionization would track the neutrino decay rate,
and neutrinos with these masses would likely lead to a ``tail'' of the
recombination epoch, rather than an episode of 2nd reionization at
lower redshift $z\sim 30$ \citep[as in][]{sarkar98}. Such a
scenario may be in conflict with existing weak constraints from CMB
polarization anisotropies~\citep{kogut03,bean03}.  In comparison, for
more massive neutrinos, reionization would be a three--step process,
as explained above: (1) the relativistic electrons produce X--ray
photons from CMB; (2) the X--ray photons photo-ionize hydrogen atoms,
and (3) the fast $\sim$keV photoelectron collisionally ionizes many
more hydrogen atoms.  Of these, (1) and (3) last only a small fraction
of the Hubble time for any choice of our model parameters.  However,
for the upper energy range we considered for the initial relativistic
electron, at redshift $z\sim 1000$, (2) will last for $\sim 10^8$yr, a
time comparable to the age of the universe at $z\sim 30$.  As a
result, for these energetic electrons, reionization will occur near
$z\sim 30$, producing a 2nd reionization at these lower redshifts.

We note further that the universe cannot be fully reionized at high
redshift in this way (Oh 2001; Venkatesan et al. 2001). While the IGM
is nearly neutral, the high energy electrons will spend about 1/3 of
their initial energy on hydrogen ionizations, and very little energy
will be deposited into heating the thermal
electrons~\citep{shull85}. Once the ionization fraction rises to
$50\%$, the fraction of the electron energy spent on ionizations drops
to only few percent, and the fraction spent on heating of the thermal
electrons rises to $90\%$~\citep{shull85}. It is therefore a
reasonable guess that the electrons will drive the ionization fraction
up to $x_e\sim 0.5$.  Note that at redshifts of $15-20$, the
recombination time is a fraction $0.5-0.3$ of the Hubble time; we
therefore expect recombinations to allow this level of partial
ionization to remain.  A universe reionized to $30(50)\%$ at redshifts
of $7<z<31(25)$, but fully reionized at $0<z<7$, has an optical depth
of $\tau_e = 0.16$ (in our adopted $\Lambda$CDM cosmology,
conservatively ignoring any contribution to the opacity from He+ or
He++ which recombine more rapidly than hydrogen). We conclude that a
scenario in which a decaying particle produces an early boost of
reionization, followed by full reionization by stars, is therefore
plausible, and can account for the {\it WMAP} optical depth.

Our scenario implies that warm dark matter (WDM) models are not ruled
out. This is not in contradiction with the results of Barkana et
al. (2001), Spergel et al. (2003) and Yoshida et al. (2003), since all
of these studies assumed that reionization was done by stars alone.
The possibility of a sterile neutrino saving the option of WDM is
rather peculiar, because the only (to our knowledge) 
natural WDM
candidate is, in fact, another sterile
neutrino~\citep{dodwid,dh01,afp01}, which is constrained to have a
mass of a few keV~\citep{abaz01,hansen02}. 
One should mention that more inventive WDM candidates certainly 
exist~\citep[see e.g.][]{baltz,boehm}.

Thus, with a minimal extension of the standard model with two sterile
neutrinos (one with a mass of about a keV, and another with a mass of
a few hundred MeV), one obtains the right abundance of dark matter,
and achieves an early boost of reionization.  WDM models may be
attractive in helping to solve some possible problems of CDM on
small--scales, although many issues remain to be clarified (see
summary in, e.g., Haiman et al. 2001).

A very important question is whether we will be able to distinguish
between early stellar reionization models, and the scenario described
above. It was shown by Kaplinghat et al. (2003) that two years of
polarization data from {\it WMAP} can distinguish between a scenario
with full reionization up to $z=6.3$, and a model with an additional
$50\%$ reionization between $z=20$ and $6.3$. Planck will do even
better in just one year, namely distinguishing between $0\%$ or $6\%$
reionization between $z=20$ and $6.3$.
For related studies, see e.g.~\citep{doro,naselsky}.
  Specifically, Haiman \& Holder
(2003) and Holder et al. (2003) showed explicit examples of different
reionization histories which produce the same optical depth
$\tau=0.16$, but can be distinguished from each other by {\it Planck}
polarization measurements at $>3\sigma$ significance.  Based on these
studies, we are confident that the model discussed above can be
constrained or confirmed within a few years. The reionization history
in the decaying neutrino model could possibly be mimicked by
ionization caused by an early generation of stars that produces
X--rays and therefore fast photoelectrons (Oh 2001; Venkatesan et
al. 2001). However, this ``degeneracy'' can be broken by examining
small angular scale Sunyaev-Zeldovich fluctuations, which should be
strong in the stellar case (Oh et al. 2003), but nonexistent in our
scenario.

There are many issues we have not touched upon in this simplified
treatment.  For example, hydrogen recombinations are significant at
high redshifts, and one would have to convolve the recombination rate
in the IGM with the neutrino decay rate, in order to compute the
actual ionization fraction as a function of redshift.  While we do not
expect such a computation to change our main conclusions, it will
allow predictions to be made about the precise evolution of the
ionization and thermal history of the IGM, as well as the resulting
spectral signatures on the CMB.

\section{Conclusions}
\label{conclusions}

In this paper, we studied the possibility of a reionization history
that consists of two completely independent steps. The first step is
provided by fast electrons from a decaying sterile neutrino. The
abundance and mass of this sterile neutrino is such that up to $50\%$
of the universe can be reionized around a redshift of
$z=25$. Subsequently, standard (population II) star formation can
complete reionization at $z\sim 6-7$. This scenario can give an
electron scattering optical depth of $\tau = 0.16$, in agreement with
{\it WMAP} observations. In such a scenario, the possibility of warm
dark matter still exists, since stars only need to complete the
reionization around redshift of $z=7$ to comply with the observed
Gunn-Peterson opacities of high redshift ($z\sim 6$) quasars.  The
model proposed here can be tested in future studies of the CMB, both
by inferring the reionization history (evolution of the ionized
fraction) from post--{\it WMAP} polarization data, and by studying the
deviation of the CMB spectrum from a pure black body shape.

\acknowledgements{%
We thank Ted Baltz, Zurab Berezhiani, Asantha Cooray, Martin Haehnelt, Peng Oh,
Subir Sarkar, and an anonymous referee for useful comments that improved this 
paper. SHH thanks the Tomalla foundation for support}%



\end{document}